\documentclass{article}
\newcommand{\be}{\begin{equation}}
\newcommand{\ee}{\end{equation}}

\begin{document}
\begin{center}{\huge The Past Earth's Rotation}
\vspace{2.5cm}
\begin{center}
{\Large Arbab I. Arbab\footnote{E-mail: arbab64@hotmail.com}}
\end{center}
\vspace{1cm} Department of Physics, Faculty of Applied Science,
Omdurman Ahlia University, Omdurman, SUDAN.
\end{center}
\vspace{2cm}

{\bf Summary}. We have proposed a model to obtain the length of the day and 
month at
any geologic time. The day is found to increase by 0.002 sec/century since 
the
seventeenth century.
The lengthening of the day is attributed entirely to the increasing
gravitational constant ($G$) with time that manifested its self in tidal 
forces.
The data obtained are consistent with those found from Palaeontology.
The length of the day 4500 million  years ago was 6 hours and the synodic
month was 56.26 present epoch days.\\

\vspace{1.5cm}
{\bf 1. INTRODUCTION}
\\
{\large There exist certain large dimensionless ratios that may be formed 
from the
fundamental constants of atomic physics and cosmology. In his large number
hypothesis (LNH) Dirac proposed that these ratios are interrelated and are
simple functions of cosmic time $t$. Thus the gravitational constant $G$, 
when
measured in units of atomic time, varies as $t^{-1}$.
The geophysical and palaeontolgical data ruled out this variation
(Blake, 1978). This is manifested in the past number of days in a year
and the length of the day derived from biological growth rhythms preserved 
in the fossil records
investigated by Wells (1963).
Based on different grounds, we have proposed a different temporal variation
for $G$ viz. $G\sim t^{(2n-1)/(1-n)}$ where $1/2<n<1$ is a constant related 
to viscosity. We emphasize
that any changes in the Earth-Moon-Sun system are entirely due to
gravitational effects that resulted from changes in $G$. Thus the
`effective' $G$ embodies all these effects.
For instance, the increase of the length of the day is generally attributed
to tidal effects by the Moon on the Earth. Some scientists in celestial
mechanics believed that lunar tides are generally responsible for the
tectonic movement of the crust. In our present approach this variation
is due to the increasing gravity that leads to change the tidal effects.
It is therefore important to associate the time parameter when recording
any information about the Earth. As $G$ changes the following parameters of
the Earth: Radius ($R$), Number of days in a year ($Y$), Length of the day 
($D$),
Earth temperature ($T$), and Moment of Inertia ($I$), will change.
In this note we will be mainly concerned with changes in $Y$ and $D$ and 
their
corresponding magnitudes.\\

\vspace{.5cm}
{\bf 2- THE MODEL }
\\

Applying the Kepler's 2nd and 3rd laws of motion to the Earth-Moon system,
neglecting the orbit eccentricity, one obtains
\begin{equation}
G^2[(M+m)^2m^3] T = 2\pi L^3\ \ ,\ \ \   G[(M+m)m^2] r = L^2
\end{equation}
and
\begin{equation}
G^2[(M+M_s)^2 M^3] Y= 2\pi N^3    \ \ , \ \  \      G[(M+M_s)M^2] R = N^2
\end{equation}
for the Earth-Sun system,
where $m$, $M$ and $M_s$ are the masses of the Moon, Earth and  Sun 
respectively;
$L$ and $N$ are their orbital angular momenta; $R$ and $r$ are the distance 
between
Moon-Earth and Earth-Sun, respectively; $Y$ and $T$ are the length of the 
year
and the sidereal month, respectively.
The synodic month ($T^s$) is related to the sidereal month by the relation
$T^s=T(1-T/Y)^{-1}$.
While the lunar tidal friction resulted in angular momentum being 
transferred
from the Earth's rotation about its axis to the orbital angular momentum of
the Moon, the orbital angular momentum of the Earth around the Sun is nearly
constant. From equation (1) and (2), one can write,
\begin{equation}
\frac{Y}{Y_0}\frac{T_0}{T}=(\frac{L}{L_0})^3\ ,
\end{equation}
a relation that is valid at any geologic time ($t$).
To fit the results obtained by Wells, one would require the gravitational
constant to vary as $G\sim t^{1.3}$ and the present age of the universe
($t_0$) to be $11\times10^9$ year. Similarly, to fit the data obtained
by Runcorn for the ratio $\frac{L}{L_0}=1.016\pm0.003$ for the Devonian, one 
requires $L$ to vary as
$L\sim t^{0.43}$. Thus one can write the time dependent (evolution)
of the above quantities as
\begin{equation}
G=G_0(\frac{t-t_0}{t_0})^{1.3}\ ,
\ Y=Y_0(\frac{t-t_0}{t_0})^{-2.6}\ ,
\end{equation}
\begin{equation}
  \ D=D_0(\frac{t-t_0}{t_0})^{2.6}\ , \ L=L_0(\frac{t-t_0}{t_0})^{0.43}\ , \ 
T=T_0(\frac{t-t_0}{t_0})^{-1.3}
\end{equation}
(in this note the subscript `0' denotes the present value of the quantity).
Table 1 shows Wells's fossil data for the length of the year in the past and
Table 2 shows our results.
In fact, the year is the same as before but the length of the day was 
shorter
than now.
Sonett {\it et al.} (1996) have shown that the length of the day 900 m.y ago 
was
19.2 hrs and the year contained 456 days. This result is indeed what we
obtain for that era.
Recently, Williams (1997) has found the length of the day for the 
Precambrian
from tidal rhythmite palaeotidal values. We would like to comment that
the values he quoted belong to different eras in our model (see Table 3).
Ksanfomality (1997) has shown that according to the principle of
isochronism all planets had an initial period of rotation between 6-8 hrs.
However, our model gives a value of 6.1 hrs.  Berry and Baker (1968) have
suggested that laminae, ridges and troughs, and bands on present day and
Cretaceous bivalve shells are growth increments of the day and month, 
respectively.
By counting the number of ridges and troughs they therefore find that
the number of days in the month in the late Cretaceous was 29.65$\pm$0.18 
and the year contained 370.3 days. Extrapolating the astronomically
determined lengthening of the day since the seventeenth century leads to
371 days.  As the Earth rotational velocity changes the Earth will adjust
its self to maintain its equilibrium (shape) that is compatible with the new
situation.
In doing so, the Earth should have experienced several geologic activities.
Accordingly, one would expect that the tectonic movements have its root in
these rotational velocity changes.\\
\vspace{.5cm}

{\bf 3- CONCLUSION }
\\

We have proposed a model for the variation of length of the day and month
based on the idea of a variable gravitational constant. This variation can
be traced back over the whole history of the Earth. All perturbations
resulting from gravitational effects are included in the coupling constant 
$G$.
The change in the Earth's parameters can be attributed to the effect of
changing gravity with time in the manner shown in this note. The influence
of gravity on the growth of these corals (biological system) is thus
manifested in tidal effects that change the length of the day and month.
These data can be inverted and used as a geological calendar.
The data we have obtained for the length of the day and month should be
checked against palaeodata that can be obtained from different disciplines.
We have in this note suggested the temporal behavior of the Earth's 
parameters
which is consistent with the hitherto known data. Further investigations
are going on in this line.\\
\vspace{.5cm}\\
{\bf ACKNOWLEDGMENTS}
\\
\\
My ideas on this subject have benefited from discussion with a
number of friends and colleagues, I am grateful to all of them.
I wish to thank the University of Khartoum for financial support.\\
\\
\vspace{.2cm} {\bf REFERENCES}
\\
\\
Arbab, A.I., 1997. {\it Gen. Relativ. Gravit. 29}, 61\\
Blake, G.M., 1978. {\it Mon. Not. R.astr. Soc.185}, 399\\
Berry, W.B. and Barker, R.M., 1968. {\it Nature 217}, 938\\
Ksanfomality, L.V., 1997. {\it Astrophys.Space Sci. 252}, 41\\
Runcorn, S.K., 1964. {\it Nature 402}, 823\\
Scrutton, C.T., 1964. {\it Palaeontology 7}, 552\\
Sonett, C.P., 1996. Kvale, E.P., Chan, M.A. and Demko,T.M., {\it Science, 
273},100 \\
Wells, J.W., 1963. {\it Nature, 197}, 948 \\
Williams, G.E., 1997. {\it Geophys. Res. Lett.24},421}\\
\newpage

\begin{table}
\caption{Data obtained  from {\it fossil corals and radiometric time}}
\vspace{1cm}
\begin{tabular}{|r|r|r|r|r|r|r|r|r|r|}
\hline
Time (million years before present) & 65 & 136 & 180 & 230 & 280 & 345 & 405 
& 500 & 600\\
\hline
solar days/year & 371 & 377 & 381 & 385 & 390 & 396 & 402 & 412 & 424\\
\hline
\end{tabular}
\end{table}

\begin{table}
\caption{Data obtained from the {\it principle of increasing gravity}}
\vspace{1cm}
\begin{tabular}{|r|r|r|r|r|r|r|r|r|r|r|r|}
\hline
Time (million years before present) & Modern & 65 & 136 &  180  & 230 & 280  
& 345  & 405 & 500 \\
\hline
solar days/synodic month & 29.53 & 29.74 & 29.97 & 30.12 & 30.28 & 30.45 & 
30.78 & 30.89 &31.22\\
\hline
solar days/sidereal month & 27.32 & 27.53 & 27.77 & 27.91 & 28.08 & 28.25 & 
28.48 & 28.69 & 29.02 \\
\hline
synodic month/year & 12.37 & 12.47 & 12.59 & 12.66 & 12.74 & 12.82 & 12.93 & 
13.04 & 13.20\\
\hline
sidereal month/year & 13.37 & 13.47 & 13.59 & 13.66 &13.74 & 13.82 & 13.93 & 
14.04 & 14.20\\
\hline
solar days/year & 365.24 & 370.9 & 377.2 & 381.2 & 385.9 & 390.6 & 396.8 & 
402.6 & 412.2\\
\hline
length of solar day (hr) & 24 & 23.6 & 23.2 & 23.0 & 22.7 & 22.4 & 22.1 & 
21.7 & 21.3\\
\hline
\end{tabular}
\end{table}
\begin{table}
\begin{tabular}{|r|r|r|r|r|r|r|r|r|r|r|r|}
\hline
Time (million years before present & 600 & 900 &1000 & 1200 & 1400 & 2000 & 
3000 & 3500 & 4500\\
\hline
solar days/synodic month & 31.58 & 32.72 & 33.11 & 33.93 & 34.79 & 37.63 & 
43.48 & 47.09 & 56.26\\
\hline
solar days/sidereal month & 29.39 & 30.53 & 30.92 & 31.75 & 32.61 & 35.46 & 
41.33 & 44.90 & 54.14 \\
\hline
synodic month/year & 13.38 & 13.94 & 14.13 & 14.54 & 14.96 & 16.35 & 19.23 & 
20.99 & 25.49\\
\hline
sidereal month/year & 14.38 & 14.94 &  15.13 & 15.54 & 15.96 & 17.35 & 20.23 
& 21.99 & 26.49\\
\hline
solar days/year & 422.6 & 456 & 467.9 & 493.2 & 520.3 & 615.4 & 835.9 & 
988.6 & 1434\\
\hline
length of solar day (hr) & 20.7 & 19.2 & 18.7 & 17.7 & 16.8 & 14.2 & 10.5 & 
8.8 & 6.1\\
\hline
\end{tabular}
\end{table}
\small
\begin{table}
\caption{Williams's data (W) in comparison with our corresponding data (A)}
\vspace{0.5cm}
\begin{tabular}{|r|r|r|r|r|r|r|r|r|r|r|}
\hline
Time (million years before present) & 620 W & 900 W& 900 W&  2500 W& 435 A& 
566 A& 1100 A& 960 A\\
\hline
solar days/synodic month & 30.5$\pm$0.5 & 31.1 & 33.0$\pm$0.4 & 32.1$\pm$1.6 
& 30.99 & 31.46 & 33.52 & 32.95 \\
\hline
solar days/sidereal month & 28.3$\pm$0.5 & 28.9 & 30.9$\pm$0.5 & 
30.0$\pm$1.8 & 28.79 & 29.26 & 31.33 & 30.76\\
\hline
synodic month/year & 13.1$\pm$0.1 & 13.47 & 14.6$\pm$0.3 & 14.5$\pm$0.5 & 
13.09 & 13.32 & 14.33 & 14.05 \\
\hline
sidereal month/year & 14.1$\pm$0.1 & 14.47 & 15.6$\pm$0.3 & 15.5$\pm$0.5 & 
14.09 & 14.32 & 15.33 & 15.05 \\
\hline
solar days/year & 400$\pm$7 & 419 & 481$\pm$4 & 465$\pm$16 & 405.6 & 419 & 
480.3 & 463.1 \\
\hline
length of solar day (hr) & 21.9$\pm$0.4 & 20.9 & 18.2$\pm$0.2 & 18.9$\pm$0.7 
& 21.6 & 20.9 & 18.2 & 18.93  \\
\hline
\end{tabular}
\end{table}
\end{document}